# HOT ACCRETION DISKS WITH ADVECTION

Xingming Chen

Department of Astronomy and Astrophysics

Göteborg University and Chalmers University of Technology

412 96 Göteborg, Sweden

Email: chen@fy.chalmers.se

## ABSTRACT

The global structure of optically thin hot accretion disks with radial advection included has been investigated. We solve the full energy conservation equation explicitly and construct the radial structure of the disk. It is found that advection is a real cooling process and that there are two solutions co-exist for a given mass accretion rate less than a critical limit. One is fully advection cooling dominated and the other is dominated by local radiative cooling. The advection dominated accretion disks are hotter than the local cooling dominated disks; they are most probably in the two-temperature regime and effects such as electron-positron pair production and annihilation may need to be considered to study the microphysics of the hot plasma. However, the global disk structure will not be much affected by the local radiative process.

# 1. INTRODUCTION

In standard accretion disk models it is assumed that the disk is axisymmetric, non self-gravitating, geometrically thin and rotating at a strictly Keplerian law. The viscosity which is responsible for both the energy generation and the angular momentum transfer is described with a phenomenological $\alpha$-model prescription (Shakura & Sunyaev 1973). Futhermore, the accretion flow is cooled by purely local radiative cooling process which balances the viscous heating exactly. The last assumption is justified if the global heat transport (radial advection) is relatively negligible. However, in optically thick slim accretion disk models in which the assumption of Keplerian rotation law is relaxed and the radial advection term in the energy equation is included explicitly, it has been revealed that advection becomes a dominant cooling process when the mass accretion rate is larger than the Eddington limit (see, e.g., Abramowicz et al 1988; Kato, Honma, & Matsumoto 1988; Chen & Taam 1993). Under that situation, the disk is both thermally and viscously stable even though the disk is radiation pressure dominated.

In optically thin hot accretion disk models, it has long been assumed that advection is not important and was neglected. Optically thin hot accretion disks have been proposed to explain the hard spectrum observed in black hole candidates such as Cyg X-1. Early work was published in mid-1970s (see Shapiro, Lightman, & Eardley 1976; Liang & Price 1977; and review by Eardley et al 1978), and these concentrate on the spectral aspects. Recent work includes Kusunose & Takahara (1988, 1989, 1990), Tritz & Tsuruta (1989), White & Lightman (1989, 1990), Björnsson & Svensson (1991a,b, 1992), and Kusunose & Mineshige (1992). These studies are focused on the microphysical processes such as electron-positron pair production and annihilation of the hot plasma (see, e.g., Svensson 1982, 1984). In all these models, global heat transport due to radial advection has not been included. Disks with advection have been studied recently by Narayan & Popham (1993), Narayan & Yi (1994) and Abramowicz et al (1995). Narayan and Popham (1993) considered accretion on to white dwarfs and got advection dominated solutions for low accretion rates when the local cooling is inefficient. Abramowicz et al (1995) obtained thermal equilibrium curves of accretion disks at a given radius with a prior assumed advection cooling term. They showed that an advection dominated, thermally stable, solution co-exists with the local cooling dominated thermally unstable hot disk solution. These results may be important not only on the theoretical aspect but also for the understanding of the observed phenomena from accretion powered sources.

Accretion disks with advection included are governed by a set of stiff differential equations. They are numerically very difficult to solve as is proven in the case of optically thick slim accretion disks surrounding black holes (see, e.g., Abramowicz et al 1988; Kato, Honma, & Matsumoto 1988; Chen & Taam 1993) and boundary layers (e.g., Narayan & Popham 1993). In the approach of Abramowicz et al (1995), a Keplerian disk is assumed and the advection term is localized by assuming the dimensionless radial differential factor, $\xi$ (see eq. [10]), to be unity. Thus local (i.e., fixed radius) solutions of the surface density can be calculated for given mass accretion rates with one algebraic equation. However, inconsistency may occur since the advection is not calculated from the global structure of the disk which has not been constructed. In this paper, we solve the global energy conservation equation explicitly and so the consistency in regard to the radial advection heat transport is insured. In the next section, we briefly describe the assumptions and formulation used in the study. In §3, we investigate the role of advection quantatively. For advection dominated solutions, an analytical formula will be derived in a self-similar approximation. Detailed numerical results on the radial structure of the accretion disks will be presented. Discussion is presented in the last section.

## 2. FORMULATION

We consider an accretion disk which is axisymmetric, non self-gravitating, optically thin and geometrically slim so that it can be described by the vertically integrated equations. In this approximation, the Keplerian rotational angular velocity is $\Omega_k = \sqrt{GM/R(R-R_G)^2}$ under the pseudo-Newtonian potential $\Phi = -GM/(R-R_G)$, where $M$ is the mass of the central object and $R_G = 2GM/c^2$ is the Schwarzschild radius (Paczyński & Wiita 1980). We assume that the angular velocity can be expressed as $\Omega = \sqrt{GM/R^3}/\omega$, thus for Keplerian disks, $\omega = 1 - R_G/R$. In a steady state, the angular momentum equation is reduced to

$$\nu\Sigma = \frac{\dot{M}}{3\pi}fg^{-1}, \qquad (1)$$

where $\dot{M}$, $\Sigma$ and $\nu$ are the mass accretion rate, the surface density, and the kinematic viscosity respectively. The correction terms are $f = 1 - 9\Omega_*/(\Omega r^2)$ and $g = -(2/3)(d\ln\Omega/d\ln R)$. Here $\Omega_* = \Omega(3R_G)$ and $r = R/R_G$. We use a standard $\alpha$-model viscosity prescription (Shakura & Sunyaev 1973):

$$\nu = \frac{2}{3}\alpha c_s H, \qquad (2)$$

where $\alpha$ is a constant, $c_s = \sqrt{p/\rho}$ is the local sound speed, and $H = c_s/\Omega_k$ is the half-thickness of the disk. Here $\rho$ and $p$ are the density and total pressure of the disk respectively, $\Sigma = 2H\rho$. In a steady state, the mass conservation equation can be written as

$$\dot{M} = -2\pi R \Sigma v_r, \tag{3}$$

where $v_r$ is the radial velocity. The steady state energy conservation equation is represented by the balance between the local viscous heating, $Q_+$, the local radiative cooling in the vertical direction, $Q_-$, and the global heat transport (radial advection), $Q_{adv}$. It is expressed as

$$Q_+ = Q_- + Q_{adv}. \tag{4}$$

The viscous heating rate per unit area is given by the standard formula

$$Q_+ = \frac{3G}{4\pi}\frac{M\dot{M}}{R^3}fg\omega^{-2}. \tag{5}$$

The advection cooling rate is taken in a form (see, e.g. Chen & Taam 1993):

$$Q_{adv} = \Sigma v_r \frac{p}{\rho}\left[\frac{4-3\beta}{\Gamma_3 - 1}\frac{d\ln T}{dR} - (4-3\beta)\frac{d\ln \Sigma}{dR}\right], \tag{6}$$

where $T$ is the mid-plane temperature of the disk, $\beta = p_g/p$ is the ratio of the gas pressure to the total pressure, $\Gamma_3 = 1 + (4-3\beta)(\gamma-1)/[\beta + 12(\gamma-1)(\beta-1)]$, and $\gamma$ is the ratio of specific heats. For optically thin disks, $p = p_g = \mathcal{R}\rho T/\mu$ and $\beta = 1$, where $\mu$ is the mean molecular weight assumed to be 0.617. We shall assume that the local radiative cooling is provided by optically thin thermal bremmstrahlung with emissivity (erg s$^{-1}$ cm$^{-2}$),

$$Q_- = Q_{brem} = 1.24 \times 10^{21} H\rho^2 T^{1/2} A, \tag{7}$$

where $A \geq 1$ is the Compton enhancement factor. Combining the above equations and the vertical hydrostatic equilibrium relations we obtain (see also Abramowicz et al 1995),

$$T = 1.39 \times 10^{13}(\alpha\Sigma)^{-1}\dot{m}r^{-3/2}fg^{-1}q^{-1}, \tag{8}$$

$$\xi\dot{m}^2 - 0.361r^{1/2}g^2q\omega^{-2}(\alpha\Sigma)\dot{m} + 3.462\times 10^{-6}A\alpha^{-2}r^2f^{-1}g(\alpha\Sigma)^3 = 0, \tag{9}$$

and

$$\xi = -\frac{1}{\gamma-1}\frac{d\ln T}{d\ln R} + \frac{d\ln \Sigma}{d\ln R}. \tag{10}$$

Here $q = 1 - R_G/R$, $\dot{m} = \dot{M}/\dot{M}_E$, and $\dot{M}_E = 4\pi GM/(c\kappa_{es})$ is defined as the Eddington rate, where $\kappa_{es} = 0.34$.

## 3. STRUCTURE OF THE DISK

For a given mass accretion rate $\dot{m}$, the radial disk structure (i.e., the surface density and temperature) can be obtained numerically from equations (8)-(10). However, the characteristic properties of the solution can be revealed under some approximations. In equation (9), one notes that the first, second and third terms correspond to the radial advection heat transport, the viscous heating and the local radiative cooling respectively. Thus by neglecting the first term, the usual purely local radiative cooling hot accretion disk solution can be recovered. Furthermore, equation (9) alone can be solved by assuming a constant $\xi$ (Abramowicz et al 1995). For a given mass accretion rate $\dot{m}$ and a fixed radius $r$, this equation is a cubic equation in $\Sigma$ and in general has three roots. They are either three real roots or one real and two complex conjugates. In the case of a postive $\xi$, which corresponds to a cooling radial advection (i.e., to a negative entropy gradient), a critical mass accretion rate exists (see Abramowicz et al 1995):

$$\dot{m}_{\max} = 1.7 \times 10^3 \alpha^2 r^{-1/2} A^{-1} f g^5 q^3 \omega^{-6} \xi^{-2}, \tag{11}$$

which decreases for a large Compton enhancement factor. For $\dot{m} < \dot{m}_{\max}$ there are one unphysical negative root of $\Sigma$ and two postive roots corresponding to the local cooling and advection cooling dominated solutions. For $\dot{m} > \dot{m}_{\max}$, however, there are one negative root and two complex conjugates. They are all unphysical. In the case of a negative $\xi$, which corresponds to a heating radial advection, there is only one physical positive solution and the other two are negative and thus unphysical.

It should be noted that, the mass of the central object does not appear in equations (8)-(10). Thus, unlike optically thick accretion disks, the temperature and surface density of hot optically thin accretion disk do not depend on the mass of the central object if they are expressed in terms of $R/R_G$ and $\dot{M}/\dot{M}_E$. This occurs because the disk is gas pressure dominated, thus $(H/R)^2 \propto T$ and no $M$ is explicitly involved. Since the heating rate, $Q_+$, scales to $M^{-1}$, it requires the cooling rates (both the local and the advection) be proportional to $M^{-1}$ also. This is the case for advection, $Q_{adv} \propto M^{-1}(H/R)^2$, and for bremmstrahlung, $Q_{brem} \propto M^{-1}(H/R)^{-1}$, but not for optically thick cooling in the radiative diffusion approximation.

## 3.1. The Role of Advection

We concentrate on the case where the radial advection is a cooling process and the mass accretion rate is less than the critical value so that two solutions can exist. To see the effect of the radial advection in each of these two solutions, the $Q_{adv}/Q_+ \sim \dot{M}$ relation is shown in Figure 1a. For simplicity, we have assumed that $A = 1$, $\omega = 1 - R_G/R$, $\xi = 1$, $\alpha = 0.1$, and $r = 10$. It is seen from the right vertical line (which corresponds to the upper-left branch line of the $\dot{M}(\Sigma)$ relation in Fig. 1b) that the disk is fully advection dominated for a wide range of mass accretion rates, that is, $Q_{adv}/Q_+ \gtrsim 0.999$. For mass accretion rates close to $\dot{m}_{\max}$, the advective and local cooling processes are equally important (see the horizotal line which corresponds to the turning point and a small section of the lower branch of the $\dot{M}(\Sigma)$ relation). However, for the rest of the solutions on the lower branch of $\dot{M}(\Sigma)$ (Fig. 1b), the advection is negligible, that is, $Q_{adv}/Q_+ \lesssim 0.001$ (see the left vertical line). This result suggests that, for mass accretion rates less than $\dot{m}_{\max}$, the usual hot accretion disk solutions can be constructed in a very good approximation by neglecting the radial advection heat transport effect, although they are thermally unstable. On the other hand, for the advection dominated solutions, the detailed local cooling process will not affect the disk structure very much. It can be easily checked that above the advection line, $Q_{adv} > Q_+$, whereas below this line, $Q_{adv} < Q_+$. Thus, for a perturbation of $\dot{M}$ (increase $\dot{M}$ but keep $\Sigma$ unchanged), or effectively an increase of temperature, the cooling rate will exceed the heating rate and so the gas cools down and the original thermal equilibrium is restored. In other words, the advection dominated solution is thermally stable.

## 3.2. Self-similar Advection Solution

The dimensionless radial differential factor, $\xi$, can be estimated from a self-similar solution of accreion disks, which can be obtained by assuming the correction terms, $f$, $g$, $q$, and $\omega$ to be constants independent of the radius of the disk. Self-similar solutions of accretion disks have been studied by various researchers (see, e.g., Begelman & Meier 1982; Spruit et al 1987; Liang 1988; Narayan & Yi 1994). Although self-similar solution has its own limitations, such as that it cannot be applied to the inner regions of the disk due to the restrictions of the boundary, it has the advantange of reducing the global radial advection to a localized quantity. The self-similar solution for advection dominated accretion disks can be obtained in the following procedure (see also Narayan & Yi 1994).

Under the assumption of the correction terms being constants, from equation (8) we have, $T\Sigma \propto r^{-3/2}$. From equation (9) by neglecting the third local cooling term, we have $\Sigma \propto r^{-1/2}$. Thus, $T \propto r^{-1}$ and

$$\xi = \frac{1}{\gamma - 1} - \frac{1}{2}, \tag{12}$$

which is independent on the disk parameters such as the mass accretion rate, the mass of the central object, and the viscous parameter $\alpha$. It gives $\xi = 1$ for $\gamma = 5/3$. It should be noted that the advection term we adopted is vertically integrated, and uncertainties are unavoidably introduced (see Abramowicz et al 1995 and Narayan & Yi 1994 for a different definition of $\xi$). However, this result can be regarded as a clear demonstration that radial advection in accretion disks is a cooling process under the self-similar assumption.

### 3.3. The Radial Disk Structure

We have solved equations (8)-(9) with an iteration technique for the local radiative cooling dominated solutions. The calculation starts at the out edge of the disk and continues inwards. By first assuming a numerical value for $\xi$, we calculate two set of solutions of $\Sigma$ and $T$ at radii $r + \Delta r/2$ and $r - \Delta r/2$ from equations (8) and (9). A new $\xi$ is then obtained from equation (10). With this new value of $\xi$, we repeat the calculation of $\Sigma$ and $T$. Finally a converged $\xi$ will be found. At the next radius, the previous $\xi$ is used as the first guess.

For the advection dominated solutions, we use a shooting method to integrate the ordinary differential equation of $\Sigma$ derived from equations (8)-(9). The initial starting point for the integration is located at a very large distance from the central object. However, solutions in regions, for example, inside $1000 R_G$, which we considered here, is not sensitive to the exact location of the starting point. The initial starting value of $\Sigma$ is obtained through iterations as stated above. For definiteness, we assume here that the disk is a Keplerian one.

We first present the relation of the dimensionless radial differential factor, $\xi$, with the radius of the disk and the mass accretion rate (Fig. 2). We see that for both type of solutions, namely, the advection dominated (solid lines) and the local cooling dominated (dotted lines), $\xi$ does not depend on the mass accretion rate very much; there is only a negligible decrease for a wide range of $\dot{m}$ (from 0.1 to 0.001). This result confirms the assumption made by Abramowicz et al (1995) that $\xi$ is a constant on the thermal equilibrium curve of $\dot{M}(\Sigma)$ for a given radius of the disk (see Figure 1b, the lower-upper branch shaped curve). It implies that the shape of the equilibrium curve will not change in more detailed models. One also notices that $\xi$ of advection dominated solution are larger than that of local cooling dominated solution and it approaches to the self-similar solution, $\xi = 1$, at larger radii of the disk. For the inner regions of the disk, $\xi$ decreases, however, due to the zero torque condition applied at the inner boundary of the disk.

The temperature of advection dominated accretion disks is much hotter than that of local cooling dominated ones and it does not depend on the mass accretion rate (Fig. 3). This reflects the fact that in this case the disk heats up to the virial temperature (Narayan & Popham 1993). On the other hand, the surface density of advection dominated accretion disks is much smaller than that of local cooling dominated disks and it depends on the mass accretion rate more sensitively(Fig. 4). For advection dominated solutions, the surface density will not depend on the local radiative process much since it is determined primarily by the radial advection and the viscous heating terms and the local cooling term has a negligible effect (see eq. [9]). The weak dependence of the temperature on the mass accretion rate and the viscous parameter, $\alpha$, is obvious from equations (8) and (9) for advection dominated disks, in which

$$\alpha \Sigma = 2.773 \dot{m} r^{-1/2} g^{-2} q^{-1} \omega^2 \xi, \tag{13}$$

and

$$T = 5.02 \times 10^{12} r^{-1} f g \omega^{-2} \xi^{-1}. \tag{14}$$

It also shows that, taking a constant $\xi = 1$ in the whole disk will give lower temperatures in the innermost regions as one then has less energy to heat the gas due to an overestimate of the advection. In the local radiative cooling case, on the other hand, we have

$$\Sigma = 3.23 \times 10^2 \dot{m}^{1/2} r^{-3/4} f^{1/2} g^{1/2} q^{1/2} \omega^{-1} A^{-1/2}, \tag{15}$$

and

$$\alpha T = 4.30 \times 10^{10} \dot{m}^{1/2} r^{-3/4} f^{1/2} g^{-3/2} q^{-3/2} \omega A^{1/2}, \tag{16}$$

which shows instead that the temperature depends on both the viscosity parameter and the mass accretion rate.

The extremely high temperature of the advection dominated accretion disks infers the presence of various microphysical processes in the hot plasma. The equations governing theses processes depend on the assumptions of the model and may include such as the electron-positron pair balance, the photon blance, the pair energy balance, and the proton energy balance. In the case of a spherical homogeneous hot plasma cloud (HPC), two parameters can be used conveniently and they fully determine the physical conditions in the HPC (see, e.g., Svensson 1982, 1984; Björnsson & Svensson 1991b). These two parameters are defined as the proton optical depth, $\tau_p$, and the global compactness, $l$:

$$\tau_p \equiv \sigma_T n_p H; \quad l \equiv \frac{L}{H} \frac{\sigma_T}{m_e c^3}, \tag{17}$$

where $\sigma_T = 6.65 \times 10^{-25} \text{cm}^2$ is the Thomoson cross section, $n_p = \rho/m_p$ is the proton number density, $m_e$ is the electron mass, $m_p$ is the proton mass, and $H$ and $L$ are the size and the luminosity of the HPC respectively. In the case of an accretion disk, Björnsson & Svensson (1991b) identified $H$ with the local scale height of the disk and replaced the global HPC compactness parameter with a local disk compactness parameter defined as

$$l = \frac{2\pi}{3} \frac{\sigma_T}{m_e c^3} H^2 Q_-. \tag{18}$$

From equations (7) and (8), we then have,

$$\tau_p = 0.2\Sigma, \tag{19}$$

and

$$l = 3.948 \times 10^4 \alpha^{-1} \Sigma \dot{m} m A f g^{-1}. \tag{20}$$

With equations (8) and (19), we can express the Comptonization parameter in $\Sigma$ and $\dot{m}$,

$$y_c = 4\Theta \, \text{Max}(\tau_p, \tau_p^2) = 1.875 \times 10^3 \alpha^{-1} \dot{m} \, \text{Max}(1, 0.2\Sigma) \, r^{-3/2} f g^{-1} q^{-1}, \tag{21}$$

where $\Theta = kT/(m_e c^2)$ is the dimensionless temperature.

The loci of $l = 10$, $\tau_p = 10^{-2}$, and $y_c = 1$ along with the thermal equilibra are shown on the $\Sigma \sim \dot{M}$ plane in Figure 5 for three different viscosity parameters, $\alpha = 0.001$, $0.01$ and $0.1$ at radius $r = 10$. Since we have expressed $\dot{M}$ in unit of $\alpha \dot{M}_\text{E}$, these loci and also the advection dominated branch do not change with $\alpha$ (see eqs. [13], [19-21]). If we assume electron-positron pair production is important for the compactness parameter, $l \gtrsim 10$ (e.g., Svensson 1984), then this condition is $\dot{M} \gtrsim 0.01 \alpha \dot{M}_\text{E}$ for advection dominated solutions, and which is higher than that of the local cooling dominated solutions. On the other hand, for both kinds of solutions, Comptonization effect becomes important only for $\dot{M} \gtrsim 0.1 \alpha \dot{M}_\text{E}$. The locus of $y_c = 1$ becomes flat (i.e., independent of $\Sigma$) when the proton optical depth, $\tau_p$, is smaller than unity.

To examine the importance of these processes at different regions of the disk, the variations of $l$, $\tau_p$, $y_c$, and $\Theta$, with respect to the radius of the disk are shown in Figure 6 for three different mass accretion rates, $\dot{m} = 0.001, 0.01$, and $0.1$, with viscosity parameter, $\alpha = 0.1$. Here the disk is advection dominated. It is seen that the disk compactness parameter does not change very much with the radius of the disk and so the electron-positron pair process is important in the entire advection dominated accretion disk for $\dot{M} \gtrsim 0.01\alpha\dot{M}_E$. The Comptonization parameter, $y_c$, however, decreases rapidly with increasing disk radius. This is due to the fact that both the temperature, $\Theta$, and the proton optical depth, $\tau_p$, decrease with radius. This result may imply that the importance of Comptonization is restricted to the inner regions of the disk. Finally, we also see that relativistic effects may be important since $\Theta \gtrsim 1$ (e.g. Svensson 1984).

Advection dominated accretion disks are most probably in the two-temperature regime since $\Theta \gtrsim 1$ (e.g. Svensson 1984). For a two-temperature accretion disk, the temperature shown here corresponds approximately to the sum of the proton (ion) temperature, $T_p$, and the electron temperature, $T_e$. Since $T_p > T_e$, we have $T \approx T_p$. The electron temperature can be determined, for example, through the proton-electron Coulomb coupling rate (see e.g., Guilbert & Stepney 1985),

$$F = 3H \left(\frac{2}{\pi}\right)^{1/2} \left(\frac{m_e}{m_p}\right) \sigma_T c (\ln \Lambda) k(T_p - T_e) \left(\frac{\rho}{m_p}\right)^2 (2z+1) \frac{1+\Theta^{1/2}}{\Theta^{3/2}}, \qquad (22)$$

where $z = n_+/n_p$, $n_+$ is the positron number density and $\ln \Lambda \approx 20$ is the Coulomb logarithm. A simple estimate can show that the electron temperature is approximately one order of magnitude smaller than the proton temperature which is in the range of $10^{11}$K to $10^{12}$K.

## 4. DISCUSSION

The global radial structure of hot optically thin accretion disks with advection included explicitly has been calculated. It is shown that advection is generally a cooling process which corresponds to a negative radial entropy gradient. For a given mass accretion rate less than a critical value, there are two solutions in which the advection can be either a dominant cooling process or negligible. Whereas for mass accretion rates above the critical value no hot optically thin accretion disk solution can exist. This result does not depend on the detailed local radiative cooling process. However, it requires that the purely local cooling solution has an approximately straight line relation of $\dot{M}(\Sigma)$ (at a fixed radius on the logarithmic plane) with a slope steeper than $d(\log \dot{M})/d(\log \Sigma) = 1/2$ or $d(\log \dot{M})/d(\log \Sigma) \leq 0$. In fact this is the case for non-pair bremsstrahlung and Comptonized soft photon cooling solutions (see, e.g., Wandel & Liang 1991, Luo & Liang 1994), and for electron-positron pair accretion disk solutions (see, e.g. Björnsson & Svensson 1992; Kusunose & Mineshige 1992).

We have used bremsstrahlung radiation mechanism for the local cooling process. At temperature exceeds $10^9$K, other radiation mechanisms can often cool the electrons much more efficiently (see e.g., Rees et al 1982). This will result the local cooling dominated lower-branch of the thermal equilibria to move in the up-left direction and the critical mass accretion rate, $\dot{m}_{\max}$, will decrease (see Fig. 5). However, these local cooling processes will not affect the advection dominated upper-branch much. This is because, as have been demonstrated in §3.1, that for advection dominated solutions, the local cooling process has a negligible effect ($Q_{adv}/Q_+ \gtrsim 0.999$). The detailed study of the microphysics of the hot plasma is beyond the scope of the present paper. However, we have suggested that the disk is in the two-temperature regime. The temperature calculated in this paper is more closely related to the proton instead of the electron temperature. A self-consistent electron temperature can be obtained by considering the detailed radiative microphysics of the hot plasma. We have also shown that complicated radiation processes are present most probably in accretion disks with large viscosity parameters and high mass accretion rates (see Fig. 5). Thus, our results are more applicable to accretion disks with small viscosity parameters and low mass accretion rates.

The advection dominated accretion disks can hardly be regarded as in a geometrically thin configuration. By considering two-dimensional (vertical and radial) effects, results obtained here will probably be changed in some way. Furthermore, the non-Keplerian effect which is obviously important due to the high local sound speed, and the transonic nature of accretion flows near the vicinity of the black hole will also play roles in determining the global disk structure. All these will be the subjects of our future investigations.

The author thanks Profs. Marek Abramowicz, Jean-Pierre Lasota and Ronald Taam for stimulating discussions and useful suggestions. Comments from them and Drs. Gunnlaugur Björnsson and Fredrik Wallinder are highly appreciated.

# FIGURE CAPTIONS

**Figure 1.**— a) The variation of $Q_{adv}/Q_+$ with respect to $\dot{m}$. Here, $A = 1$, $\omega = 1 - R_G/R$, $\xi = 1$, $\alpha = 0.1$, and $r = 10$. b) The $\dot{m}(\Sigma)$ relation. Note that on the upper branch, the advection is fully dominated whereas on the lower branch, the advection is negligible.

**Figure 2.**— The variation of the dimensionless advection factor $\xi$ with respect to the radius of the disk and the mass accretion rate. The solid and dotted lines are for advection and local cooling dominated respectively. Here, $\alpha = 0.1$ and $\dot{m} = 0.001 - 0.1$. Note that $\xi$ does not depend on the mass accretion rate very much. There is only a negligible decrease for a wide range of $\dot{m}$ (0.1 to 0.001).

**Figure 3.**— The radial temperature structure of hot optically thin accretion disks. The dotted lines from top down represent the local cooling dominated solutions for $\dot{m} = 0.1, 0.01$ and $0.001$ respectively. The solid lines represent the advection dominated solutions which do not depend on the mass accretion rate.

**Figure 4.**— The radial surface density distribution of hot optically thin accretion disks. The dotted lines from top down represent the local cooling dominated solutions for $\dot{m} = 0.1, 0.01$ and $0.001$ respectively. The solid lines from top down represent the advection dominated solutions for $\dot{m} = 0.1, 0.01$ and $0.001$ respectively.

**Figure 5.**— The loci of $l = 10$, $\tau_p = 10^{-2}$, and $y_c = 1$ along with the thermal equilibra. Here, $r = 10$ is assumed. Note that pair production is important for $\dot{M} \gtrsim 0.01\alpha\dot{M}_E$ (advection dominated solutions); and Comptonization effect becomes important for $\dot{M} \gtrsim 0.1\alpha\dot{M}_E$ (both kinds of solutions).

**Figure 6.**— The variations of $l$, $\tau_p$, $y_c$, and $\Theta$, with respect to the radius of the disk. Here the disk is advection dominated and $\alpha = 0.1$. The solid, dotted, and dashed lines represent $\dot{m} = 0.1$, $0.01$, and $0.001$ respectively. Note that the disk compactness parameter does not change very much with the radius while the Comptonization parameter decreases rapidly.